\documentclass[]{aa}

\def\rmit#1{{\it #1}}            
\def\specchar#1{{\sc #1}}

\def\NaIDtwo{\mbox{Na\,\specchar{i}\,\,D$_2$}}
\def\NaIDone{\mbox{Na\,\specchar{i}\,\,D$_1$}}

\def\HeI{\mbox{He\,\specchar{i}}}
\def\Ha{\mbox{H\,$\alpha$}}

\def\CaII{\mbox{Ca\,\specchar{ii}}} 

\def\arcsec{\hbox{$^{\prime\prime}$}}

\def\eg{\rmit{e.g.}}

\def\arcsec{\hbox{$^{\prime\prime}$}}

\usepackage{graphicx}
\usepackage{xcolor}

\usepackage{txfonts}

\usepackage[colorlinks=false]{hyperref}

\usepackage{soul} 

\begin{document}

\title{Detection of resonant nodes in a pore chromosphere}

   \author{T. Felipe,
          \inst{1,2}\fnmsep\thanks{tobias@iac.es}
          H. Kumar,
          \inst{1,2}
           \and
          E. G. Broock\inst{1,2}
          }

   \institute{Instituto de Astrof\'{\i}sica de Canarias 
              38205 C/ V\'{\i}a L{\'a}ctea, s/n, La Laguna, Tenerife, Spain
         \and
             Departamento de Astrof\'{\i}sica, Universidad de La Laguna
             38205, La Laguna, Tenerife, Spain 
            }

   \date{Received ; accepted }

 \titlerunning{Resonant nodes in a pore}
  \authorrunning{Felipe et al.}

  \abstract
   {Active region atmospheres host many oscillatory phenomena. The chromosphere is delimited by steep temperature gradients at the photosphere and transition region, where magnetoacoustic waves are trapped and can form standing oscillations within this resonant cavity.}
   {We aim to detect the signature of the resonant nodes of standing waves, which are expected to produce sudden jumps in the oscillatory phase and power dips.}
   {Spectroscopic temporal series of \Ha\ in a pore were acquired with the Swedish Solar Telescope. The velocity and temperature fluctuations at multiple atmospheric heights were inferred from the analysis of the intensity at many spectral positions along the line wings. Wavelet analysis was employed to characterize the phase differences and the power at different heights.}
   {The phase shift between velocity and temperature shows a $\pm90^{\circ}$ value, which is consistent with standing oscillations. Robust evidence of the presence of a nodal layer in the temperature at around the height probed by the intensity at $\Ha\pm0.30$ \AA\ is found, such as the detection of 180$^{\circ}$ jumps in the phase of the temperature oscillations and remarkable power dips at the same atmospheric layer. The exact height of the resonant nodes depends on the spatial location and time. We generally find a mixture of standing and propagating waves. This is consistent with a leaky resonator where waves are partially reflected at the transition region, while some of them are able to propagate into the corona. }
   {For the first time, we report the detection and characterization of resonant nodes in the solar chromosphere. This result provides strong observational support for the chromospheric resonant cavity model and paves the way for the development of new seismological techniques to investigate the structure of active region chromospheres.}

   \keywords{Solar chromosphere --- Sunspots --- Solar atmosphere --- Solar oscillations}

   \maketitle
%

\section{Introduction}\label{sec:intro}

The solar atmosphere is permeated by numerous magnetohydrodynamic wave modes. They are a key ingredient to understand the energetic balance of the upper atmospheric layers and also provide valuable diagnostics of the atmospheric structure. In sunspots, where strong magnetic fields are found, these waves have generally been detected as slow magnetoacoustic modes that propagate along magnetic field lines \citep{Bel+Leroy1977, Lites1984, Zhugzhda+Dzhalilov1984c, Bloomfield+etal2007b}. Oscillations have been reported over a broad range of atmospheric heights, from the photosphere to the corona. At the chromosphere, they are dominated by oscillations with periods in the three-minute band \citep{Lites1986, Centeno+etal2006}. These waves have been widely studied through observations, analytical studies, and numerical simulations since they are one of the fundamental aspects of the dynamics in magnetized solar plasma. The interested reader is referred to the numerous comprehensive reviews available in the literature for a more detailed overview of this topic \citep[\eg,][]{Bogdan+Judge2006, Khomenko+Collados2015, Jess+etal2023}.

\begin{figure*}[ht]  
\centering
\includegraphics[width=0.95\textwidth]{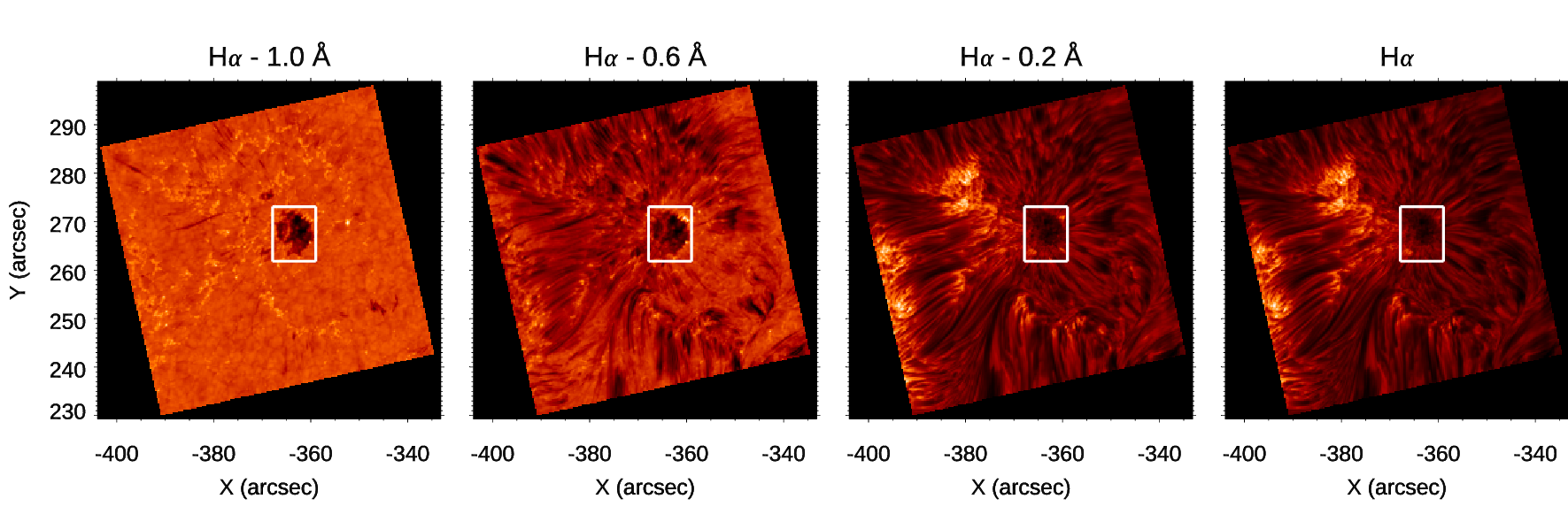}
\caption{Field of view of CRISP \Ha\ observations at different wavelengths from the far blue wing (left panel) to the core of the line (right panel). The wavelength is indicated at the top of the panel. The white squares delimit the region illustrated in Fig. \ref{fig:dphase_maps}.  } \label{fig:mapa}
\end{figure*}

Despite decades of research, the mechanisms producing the enhancement of the three-minute chromospheric oscillations are still under debate. The most favored interpretation is based on the acoustic cut-off frequency, which filters the photospheric waves by allowing the propagation of only those with frequencies above approximately 5 mHz, while low frequency waves become evanescent \citep{Fleck+Schmitz1991, Centeno+etal2006, Felipe+etal2010b}. Numerous phase shift measurements between the oscillatory signals at different atmospheric layers support this scenario \citep{Centeno+etal2009, KrishnaPrasad+etal2017}. 

The complexity of the chromospheric power spectrum, which contains numerous peaks extending beyond the dominant three-minute band, suggests that additional physical processes may contribute to the observed oscillatory behavior \citep{Jess+etal2020, Sych+etal2024, Sych+Yan2025}. This alternative explanation proposes the existence of a chromospheric resonant cavity bounded by the steep temperature gradients at the photosphere and transition region \citep{Zhugzhda+Locans1981, Gurman+Leibacher1984, Zhugzhda2008}. In this scenario, slow magnetoacoustic waves are trapped by partial reflections at those boundaries that amplify discrete frequencies and form standing modes \citep{Botha+etal2011, Snow+etal2015, Felipe2019}. The detection of a high-frequency power peak around 20 mHz has been identified as a signature of this chromospheric resonator \citep{Jess+etal2020}, although similar spectral signatures can also arise from nonlinear wave evolution without the need of a reflecting boundary \citep{Felipe2021}. However, independent analysis of the phase relations between velocity and temperature (intensity) fluctuations are also consistent with chromospheric resonances \citep{Felipe+etal2020, Sangal+etal2026}.

A definitive observational proof of standing waves is the detection of resonant nodes, where the amplitude of the fluctuations approaches a minimum while the oscillatory phase exhibits an abrupt 180$^{\circ}$ jump across the node \citep{Fleck+Deubner1989}. These nodal signatures constitute an unambiguous means of distinguishing standing waves from freely propagating disturbances, making the identification of resonant nodes a critical observational test of the chromospheric resonance hypothesis. \citet{Felipe+etal2025} found indications of the presence of resonant nodes and dynamic changes in their atmospheric height from the analysis of chromospheric oscillations with umbral flashes.

In this work, we characterize for the first time the sudden transitions in the oscillatory phase and the power dips produced by chromospheric resonant nodes. The confirmation of these signatures provides strong observational support for resonance as a key mechanism governing wave dynamics in sunspot atmospheres, helping to resolve the long-standing debate regarding the origin of chromospheric three-minute oscillations and the relative importance of resonant trapping versus wave propagation. In Sect. \ref{sect:observations}, we describe the observational data. Section \ref{sect:results} contains the results, including the wavelet analysis performed to measure the oscillatory phase and detect the resonant nodes. Finally, Sect. \ref{sect:conclusions} discusses the results and summarize the conclusions.

\section{Observations} \label{sect:observations}

The leading pore of active region NOAA 12848 was observed on 21 July 2021 with the Swedish Solar Telescope \citep[SST; ][]{Scharmer+etal2003}. At the time of the observations, the target was positioned at helioprojective coordinates ($x = -364$\arcsec, $y = 271$\arcsec). Spectral imaging in the \CaII\ 8542 \AA\ and \Ha\ lines was carried out using the CRisp Imaging SpectroPolarimeter \citep[CRISP; ][]{Scharmer2006, Scharmer+etal2008}. The present work concentrates on the \Ha\ observations, specifically on a continuous time series covering the interval from 08:04 UT to 09:05 UT with a total of 130 time steps.

CRISP provides quasi-monochromatic imaging at predefined wavelength offsets with a spatial scale of $0.\arcsec0592$ per pixel. The \Ha\ profile was sampled at 17 spectral positions within the range $\Delta\lambda=\pm1000$ m\AA. A finer wavelength spacing of 100 m\AA\ was used between $-600$ and $+600$ m\AA, while the outer wings of the line were sampled every 200 m\AA. Completing a full spectral scan of both observed lines required 28.0 s. Image restoration and reduction were performed with the SSTRED pipeline \citep{Lofdahl+etal2021}, which incorporates the multi-object multi-frame blind deconvolution method \citep[MOMFBD; ][]{Lofdahl2002,vanNoort+etal2005} to mitigate atmospheric distortions and improve image quality.

Figure \ref{fig:mapa} illustrates the field of view of the observations at some selected wavelengths. The \Ha\ line probes the solar atmosphere from the upper photosphere (far wings of the line, left panel) to the mid-to-upper chromosphere (line core, right panel) \citep{Leenaarts+etal2012}. We have evaluated the Doppler velocity and the line intensity at many spectral positions along the line profile, using the average and difference of intensity between both wings as diagnostics \citep[\eg,][]{Watanabe+etal2011}. The intensity at the spectral position $\Delta\lambda$ was computed for all spatial locations $(x,y)$ and all the temporal steps $t$ of the observation as $I_{\rm av}(x,y,t,\Delta\lambda)=(I_{\rm +\Delta\lambda}(x,y,t)+I_{\rm -\Delta\lambda}(x,y,t))/2$, where $I_{\rm \pm\Delta\lambda}$ indicate the intensity in the blue ($-$) and red ($+$) wing of the spectral line. This average of the intensity at both sides of the line was computed to remove to first order the contribution of the Doppler shift to the intensity fluctuations. As a proxy for the Doppler velocity we employed $V_{\rm diff}(x,y,t,\Delta\lambda)=(I_{\rm \Delta\lambda}(x,y,t)-I_{\rm -\Delta\lambda}(x,y,t))/(I_{\rm +\Delta\lambda}(x,y,t)+I_{\rm -\Delta\lambda}(x,y,t))$. Under this definition, a positive (negative) velocity corresponds to a downflow (upflow). The quantities $V_{\rm diff}$ and $I_{\rm av}$ were computed at $\Delta\lambda$ in the range between 0.05 and 1.00 \AA\ in steps of 0.05 \AA\ for a total of 20 measurements probing the fluctuations at multiple heights.

\begin{figure}[ht] 
\centering
\includegraphics[width=0.45\textwidth]{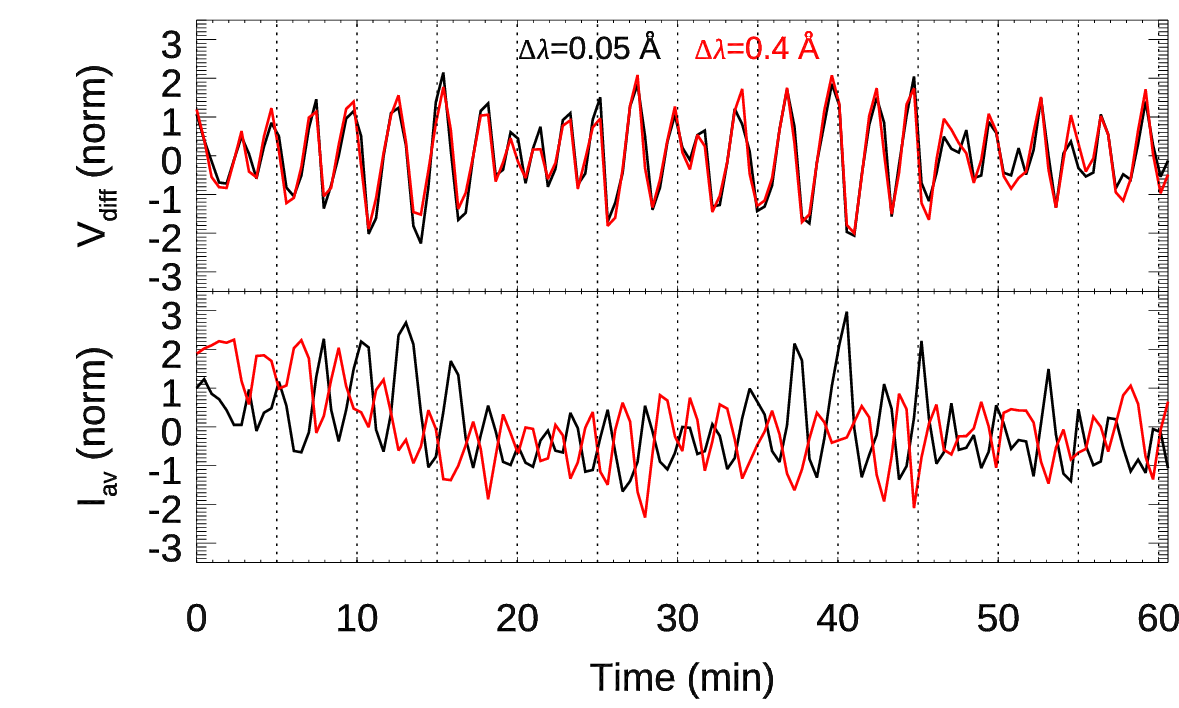}
\caption{Temporal evolution of the velocity ($V_{\rm diff}$, top panel) and temperature ($I_{\rm av}$, bottom panel) at a randomly selected pore location. Fluctuations are shown at two different atmospheric layers, corresponding to $\Delta\lambda=0.05$ \AA\ (black lines) and  $\Delta\lambda=0.40$ \AA\ (red lines).} \label{fig:fluctuations}
\end{figure}

\section{Results} \label{sect:results}

\subsection{Velocity and temperature fluctuations} 

For each spatial position and atmospheric height (given by $\Delta\lambda$), the zero values of $V_{\rm diff}$ and $I_{\rm av}$ have been set at the average over the whole temporal series, and the fluctuations have been normalized to the standard deviation. While $V_{\rm diff}$ provides a reliable estimation of the phase of the Doppler velocity, the interpretation of $I_{\rm av}$ is more complex. According to \citet{Leenaarts+etal2012}, in the quiet Sun at heights below 1 Mm, the \Ha\ line is sensitive to the temperature, whereas at higher layers it is mainly sensitive to density. Due to the lower temperature of active regions, we would expect the sensitivity of \Ha\ to temperature to be extended to higher layers in the observed pore. Also, assuming adiabatic waves, the temperature and density fluctuations are in phase. All in all, we consider that $I_{\rm av}$ provides an accurate characterization of the phase of temperature fluctuations. In the following, we will refer to $V_{\rm diff}$ as velocity and to $I_{\rm av}$ as temperature.

Figure \ref{fig:fluctuations} shows the temporal evolution of the fluctuations in velocity and temperature at a selected location within the pore. Both signals exhibit a periodicity around 3 min, as expected for chromospheric waves. The velocities measured at $\Delta\lambda=0.05$ and $\Delta\lambda=0.40$ \AA\ barely show any differences. They fluctuate in phase during the whole temporal series, even though they probe different atmospheric layers. This result is consistent with the presence of standing oscillations. In contrast, the temperature inferred at both atmospheric heights is fluctuating in opposite phase. The temperature enhancements at $\Delta\lambda=0.05$ \AA\ are accompanied by temperature minima at $\Delta\lambda=0.40$ \AA\ and vice versa. This is also consistent with the behavior of standing oscillations when a resonant node is present between the formation height of those signals. Oscillations at opposite sides of a temperature resonant node fluctuate with a 180$^\circ$ phase difference.   

\begin{figure}[ht] 
\centering
\includegraphics[width=0.45\textwidth]{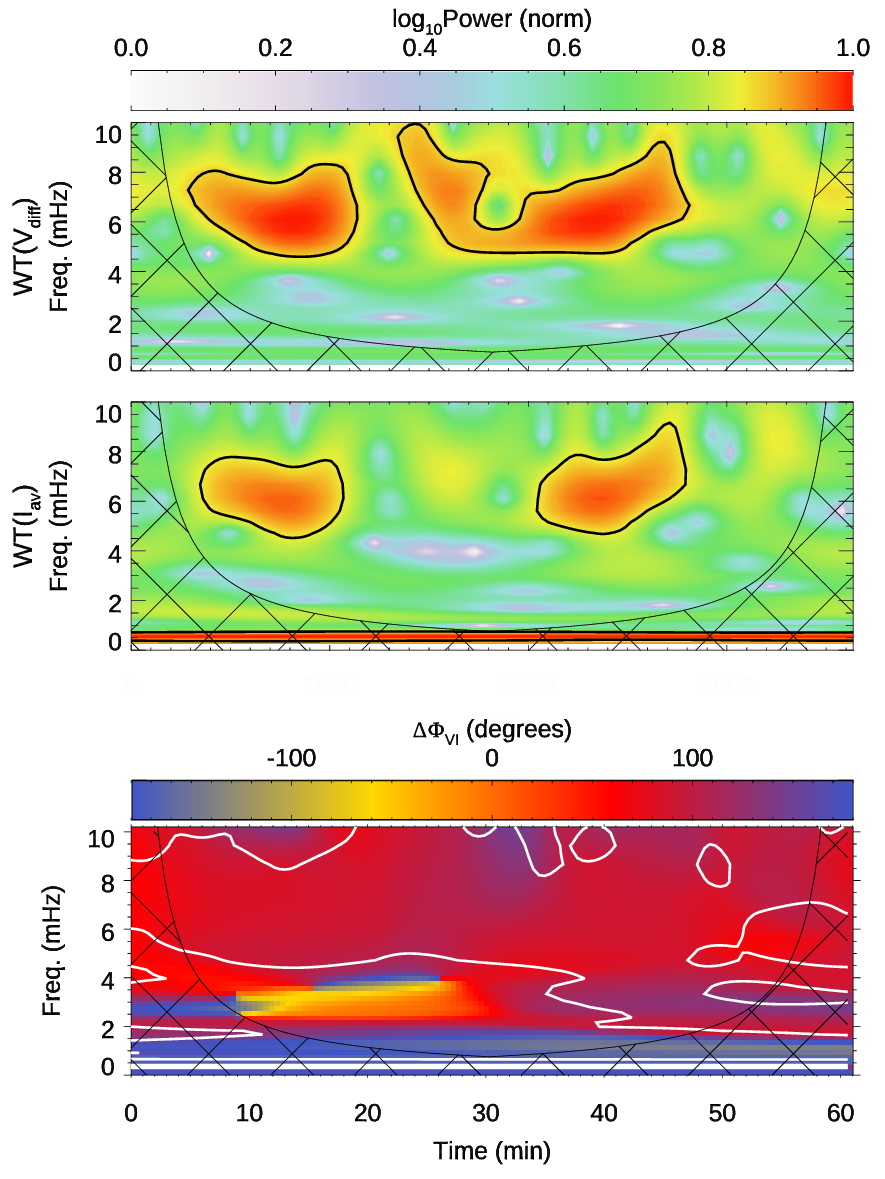}
\caption{Wavelet analysis at of the signals at $\Delta\lambda=0.05$ \AA\ illustrated in Fig. \ref{fig:fluctuations}. Top panel: Wavelet power of the velocity. Middle panel: Wavelet power of the temperature. Bottom panel: Phase difference between velocity and temperature. The gridded region indicates the parts of the spectra outside the cone of influence. Solid black/white lines delimit the 95\% confidence level.} \label{fig:wavelet}
\end{figure}

A visual examination of the phase relations between velocity and temperature shows that at $\Delta\lambda=0.05$ \AA\ the temperature signal is lagging the velocity by approximately a quarter period (+90$^\circ$). At the lower atmospheric height probed by $\Delta\lambda=0.40$ \AA, the velocity maintains the same phase from higher atmospheric heights while the temperature fluctuates in anti-phase, meaning that the phase difference between velocity and temperature is -90$^\circ$. These quarter-period phase differences ($\pm90^\circ$) are also consistent with standing waves \citep[\eg,][]{Deubner+etal1990}, and allow us to discriminate them from the adiabatic propagation of linear waves, where velocity and temperature oscillations are in-phase (0$^\circ$) or anti-phase (180$^\circ$). Radiative heating and cooling can modify the phase relations between velocity and intensity, and a $90^\circ$ phase difference could also be interpreted as a signature of propagating waves \citep{Chae+etal2023}. However, this effect cannot explain the sudden 180$^\circ$ phase jump between temperature fluctuations at closely separated atmospheric heights, which must be produced by a resonant node. Thus, our results are consistent with predominantly adiabatic waves trapped in a chromospheric cavity.

\subsection{Wavelet analysis}

Wavelet analysis \citep{Torrence+Compo1998} has been employed to explore the wave content of the time series. This technique has been widely used to study oscillations in the solar atmosphere \citep[\eg,][]{Bloomfield+etal2004, LohnerBottcher+BelloGonzalez2015,GuevaraGomez+etal2021, Jafarzadeh+etal2026}. For each spatial location within the white rectangle in Fig. \ref{fig:mapa}, we compute the Morlet wavelet transform of the temporal series $V_{\rm diff}(t)$ and $I_{\rm av}(t)$. This transform decomposes the time series in time and frequency domains, allowing the characterization of the dominant frequency modes and their temporal evolution. Also, cross-wavelet analysis can be used to measure the phase difference between two signals.  

Figure \ref{fig:wavelet} illustrates the wavelet power of the velocity and temperature, and the phase difference between both signals at the same location illustrated in Fig. \ref{fig:fluctuations}, for the case $\Delta\lambda=0.05$ \AA. Both velocity and temperature fluctuations are dominated by oscillations in the three-minute band, with power peaks around 6.1 mHz and strong power above the 95\% confidence level in the frequency range between 5 and 8 mHz. High-power oscillations in the three-minute band do not span the whole temporal series. Instead, they are mainly detected in two windows centered at around $t=10$ min and $t=40$ min, coinciding with the times with strong amplitude oscillations in Fig. \ref{fig:fluctuations}. The phase difference between velocity and temperature is around $90^{\circ}$ for all the regions of the time-frequency domain where the measurement is above the 95\% confidence, with some small departures to higher phase differences. This positive phase difference indicates that the temperature signal is lagging the velocity fluctuations, with a quarter-period delay that is consistent with standing waves. Notably, the confidence of the phase shift in the three-minute band is high during almost all the temporal series, not only at the times where strong oscillations were found.

\begin{figure*}[ht] 
\sidecaption
\includegraphics[width=12cm]{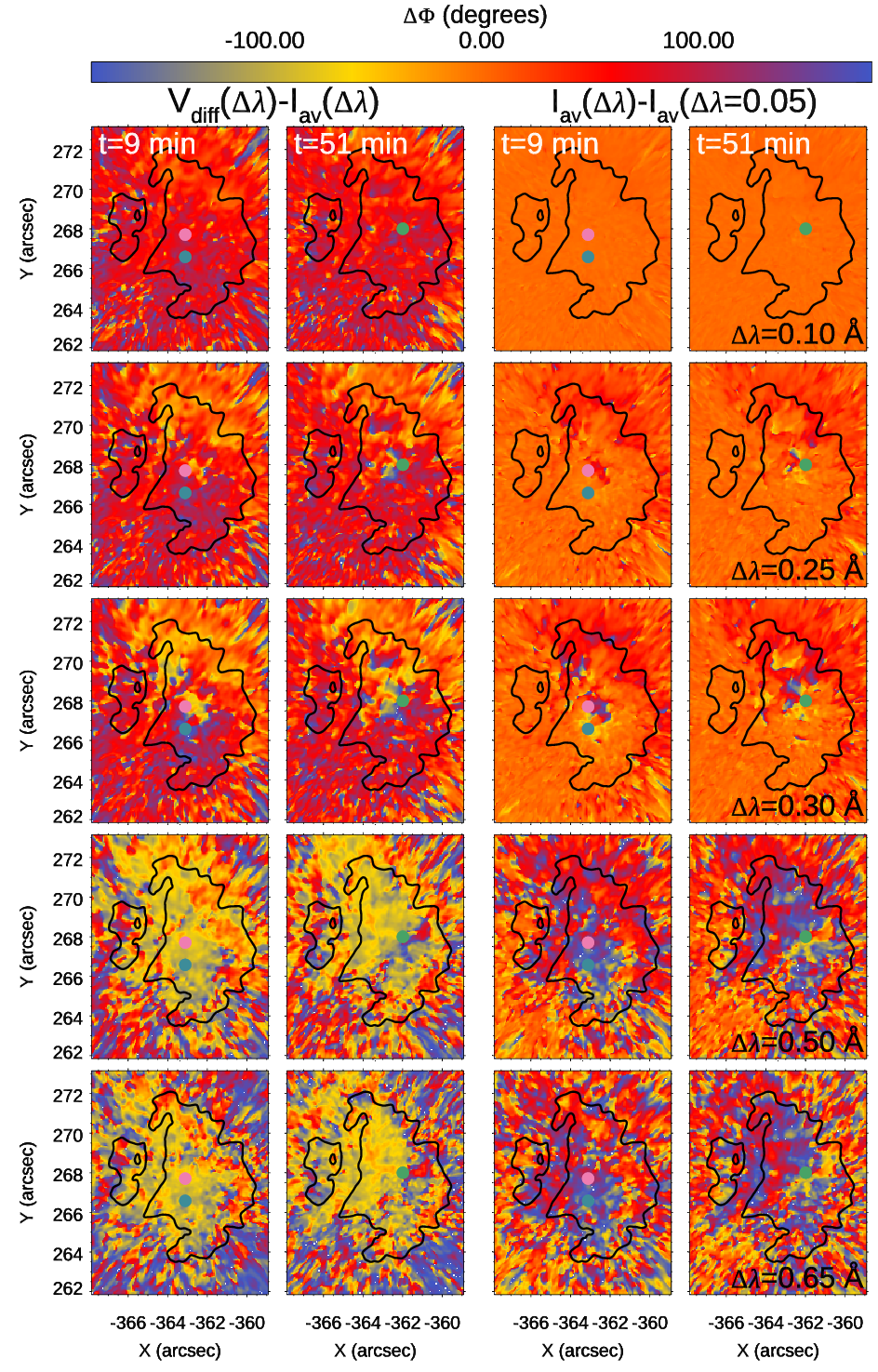}
\caption{Phase difference maps at 6.1 mHz for several times and heights. Two left-most columns: Phase difference between $V_{\rm diff}$ (velocity) and $I_{\rm av}$ (temperature). Two right-most columns: Phase difference between  $I_{\rm av}$ at different $\Delta\lambda$ and $I_{\rm av}$ at the highest probed layer $\Delta\lambda=0.05$ \AA. Rows correspond to different heights given by the $\Delta\lambda$ value indicated in the bottom-right corner of the right column, from higher layers (top row) to deeper layers (bottom row). Each pair of columns illustrates the phase differences at two different time steps (t=9 min and t=51 min). Black lines show contours of constant intensity in the average intensity, delimiting the pore region. Color dots indicate the locations and times of the phase differences plotted in Fig. \ref{fig:dphase_height}.  } \label{fig:dphase_maps}
\end{figure*}

\begin{figure*}[ht]  
\centering
\includegraphics[width=0.85\textwidth]{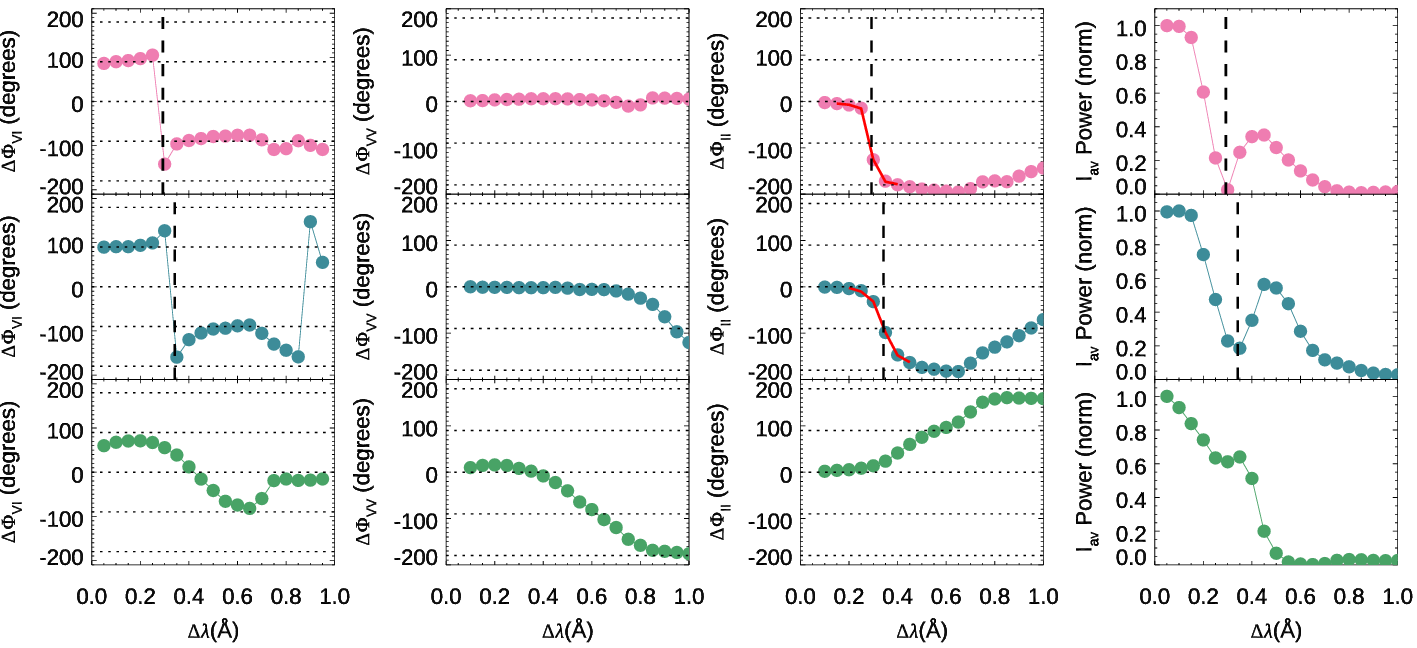}
\caption{Phase difference at 6.1 mHz as a function of atmospheric height. Each row corresponds to a different time and/or spatial position, as indicated by the corresponding dot colors in Fig. \ref{fig:dphase_maps}. First column: Phase difference between $V_{\rm diff}$ (velocity) and $I_{\rm av}$ (temperature). Second column: Phase difference between  $V_{\rm diff}$ at different $\Delta\lambda$ and $V_{\rm dff}$ at the highest probed layer ($\Delta\lambda=0.05$ \AA). Third column: Phase difference between  $I_{\rm av}$ at different $\Delta\lambda$ and $I_{\rm av}$ at the highest probed layer ($\Delta\lambda=0.05$ \AA). Red lines show the results of the arctan function fitting the phase transition. Vertical dashed lines illustrate the central height of the transition determined from the fitting. Horizontal dotted lines are added as visual aids at -180$^{\circ}$, -90$^{\circ}$, 0$^{\circ}$, 90$^{\circ}$, and 180$^{\circ}$. Fourth column: Wavelet power of the temperature at 6.1 mHz, normalized to the maximum value.} \label{fig:dphase_height}
\end{figure*}

\subsection{Wavelet phase differences and node identification}

Figure \ref{fig:dphase_maps} illustrates phase difference maps sampling two different time steps, one near the beginning of the observed series and another near the end, and several atmospheric heights from the fluctuations measured at distinct $\Delta\lambda$ along the wings of \Ha. The two left-most columns show the phase difference between velocity and temperature at 6.1 mHz, where the power of their fluctuations is maximum. In the following, we will exclusively focus on waves with this frequency. At higher layers ($\Delta\lambda=0.10$ \AA), most of the field of view is dominated by fluctuations with a 90$^{\circ}$ (red color) phase difference between velocity and temperature. In contrast, the lower atmospheric heights probed by the wings of \Ha\ at $\Delta\lambda\ge0.5$ \AA\ exhibit a phase shift around -90$^{\circ}$ (yellow color) in most of the pore region, including the light bridge, and some mixed phase differences outside the pore. Both -90$^{\circ}$ and 90$^{\circ}$ V-T phase differences are consistent with standing oscillations, and point to the presence of a resonant node between the atmospheric layers probed by the core of \Ha\ and those probed by the wings. 

The examination of the phase difference between the temperature oscillations measured at several $\Delta\lambda$ along the \Ha\ line and the highest layer probed in our analysis ($\Delta\lambda=0.05$ \AA) also reveals the existence of a resonant node, confirming it as a temperature resonant node. The temperature fluctuations at higher layers are oscillating in phase, but some departures in the phase are noticeable at $\Delta\lambda=0.25$ \AA, mostly near the center of the pore. At around $\Delta\lambda=0.50$ \AA, a significant region of the pore area is oscillating with opposite phase with respect to the upper layers.

The 180$^{\circ}$ jumps in the phase probing the presence of a resonant node are better illustrated in Fig. \ref{fig:dphase_height} for some selected spatial locations and times. The top and middle panels show the V-I, V-V, and I-I phase differences ($\Delta\phi_{\rm VI}$, $\Delta\phi_{\rm VV}$ and $\Delta\phi_{\rm II}$) and the temperature power at the locations indicated by pink and blue dots, respectively, in the first and third columns from Fig. \ref{fig:dphase_maps}. {In the phase difference plots involving temperature}, there is a sharp transition in the phase differences around $\Delta\lambda=0.30$ \AA. At both sides of this jump, the phase of the oscillations is approximately constant, with $\Delta\phi_{\rm VI}=90^{\circ}$ at higher layers and $\Delta\phi_{\rm VI}=-90^{\circ}$ at deeper layers. The temperature fluctuations at $\Delta\lambda=0.05$ \AA\ oscillate in phase with the temperature at $\Delta\lambda<0.25$ \AA, while just below this layer a sudden phase jump of 180$^{\circ}$ is found. It confirms that the phase jump detected in $\Delta\phi_{\rm VI}$ is produced by a change in the phase of the temperature fluctuations due to the presence of a temperature resonant node, whereas the velocity fluctuations are mostly in phase at all these atmospheric layers (see second column from Fig. \ref{fig:dphase_height}). Another signature of the nodal layer is found in the variation of the power of the temperature fluctuations with height (right column from Fig. \ref{fig:dphase_height}). At the same layers where the phase jump is found, temperature oscillations exhibit a remarkable reduction in their power.

\begin{figure*}[ht]  
\centering
\includegraphics[width=0.85\textwidth]{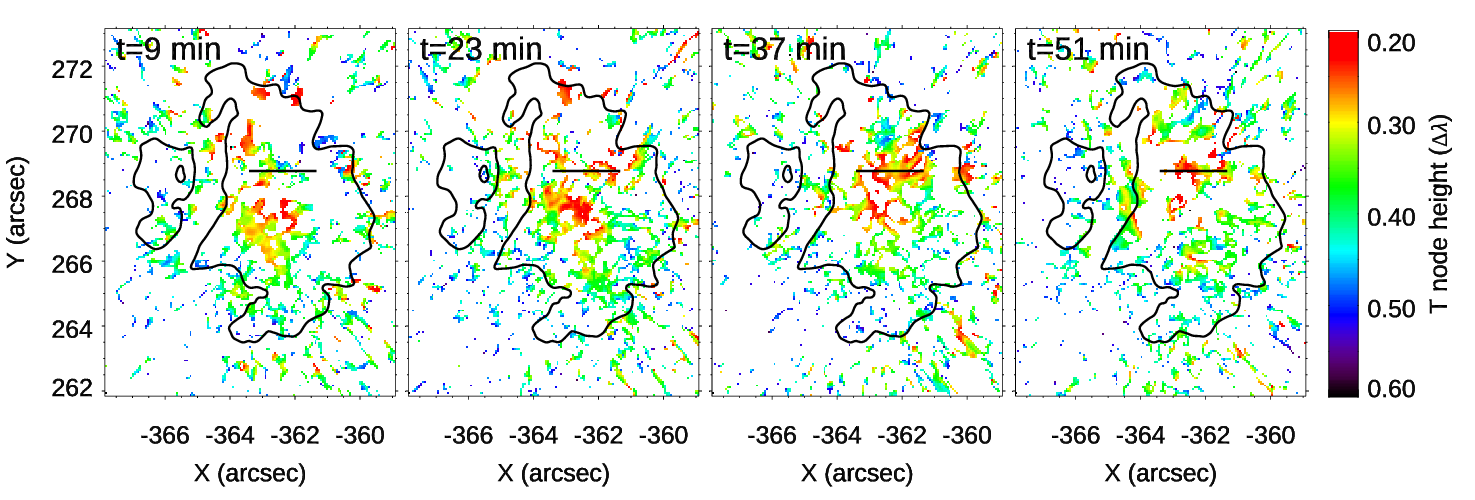}
\caption{Maps of the temperature resonant node height at 6.1 mHz at four different time steps (time shown at the top-left of each panel). In the white areas the node location was not identified. Black lines show contours of constant intensity in the average continuum intensity, delimiting the pore region. The horizontal black line indicates the location plotted in Fig. \ref{fig:nodos_time}.} \label{fig:nodos_maps}
\end{figure*}

For $\Delta\lambda>0.65$ \AA, $\Delta\phi_{\rm VI}$ starts to depart from -90$^{\circ}$, whereas $\Delta\phi_{\rm II}$ and $\Delta\phi_{\rm VV}$ also exhibit progressive changes with height. This points to the presence of propagating waves. At those deep layers, waves are not pure standing oscillations, but there is net propagation instead. Even though the oscillatory power at those heights is low (see the right column of Fig. \ref{fig:dphase_height}), the confidence level of the cross-wavelet between the temperature signal at deep layers ($\Delta\lambda > 0.65$ \AA) and that at $\Delta\lambda = 0.05$ \AA\ is above 0.75 in all cases, demonstrating that the indication of partially propagating waves is a robust result. The V-V phase differences (second column of Fig. \ref{fig:dphase_height}) also show smooth variations at deep layers, but in this case, the confidence level is low, being below 0.3 in many cases.

In summary, at these locations we find the presence of standing oscillations at layers above $\Delta\lambda=0.65$ \AA. The velocity is in phase for all the atmospheric layers probed by \Ha\ above that height, while the temperature shows a 180$^{\circ}$ jump due to the presence of a resonant node.

We have quantified the height, as given by $\Delta\lambda$, of the temperature resonant nodes at 6.1 mHz by fitting the approximately 180$^{\circ}$ jump in the phase difference $\Delta\phi_{\rm II}$ as a function of $\Delta\lambda$ to an arctan function. The node height is assigned to the $\Delta\lambda$ value of the center of the transition. The arctan function provides a smooth, mathematically continuous transition between the two sides of the phase jump, mimicking the observed changes. The addition of a scaling factor allows us to control the steepness of the transition, thereby fitting the measurements and accurately determining the location of the transition. This approach has been applied to all the spatial locations from the field of view illustrated in Fig. \ref{fig:dphase_maps} and for all the time steps (inside the cone of influence) where a sharp enough transition was found. The phase change is considered sufficiently sharp when the phase difference jump goes from zero to at least $\pm160^{\circ}$ within the layers where $\Delta\lambda\leq 0.6$ \AA. The red lines in top and middle rows from the third column of Fig. \ref{fig:dphase_height} illustrate the fit of those phase jumps, while the vertical dashed lines indicate the center of the transition that was selected as the node location. The height of the detected transition centers remarkably agrees with the height where a minimum in the temperature power is found (right column from Fig. \ref{fig:dphase_height}), providing compelling evidence that this measurement arises from the presence of a nodal layer in the temperature.

In contrast, bottom panels correspond to a time and location where the phase progressively changes, probably due to the coexistence of propagating waves instead of pure standing oscillations. The increase in $\Delta\phi_{\rm II}$ with depth indicates that the temperature signal at the highest layer ($\Delta\lambda=0.05$ \AA) is delayed with respect to deeper heights, with a larger delay as deeper layers are considered. That is, it is indicative of net upward wave propagation co-existing with standing fluctuations. A dip in the power of the temperature fluctuations is also found around $\Delta\lambda=0.30$ \AA\ (bottom right panel from Fig. \ref{fig:dphase_height}), but it is not as prominent as those identified in the other two cases.

An examination of $\Delta\phi_{\rm II}$ and $\Delta\phi_{\rm VI}$ across different times and locations reveals substantial variability in the oscillatory behavior of the waves, ranging from clear standing oscillations (similar to those shown in the top and middle panels of Fig.~\ref{fig:dphase_height}) to predominantly propagating waves. Intermediate cases, exhibiting a combination of standing and propagating wave characteristics, are also observed. The height of the nodes has only been determined for the cases where oscillations are dominated by standing waves.

\begin{figure}[ht] 
\centering
\includegraphics[width=0.45\textwidth]{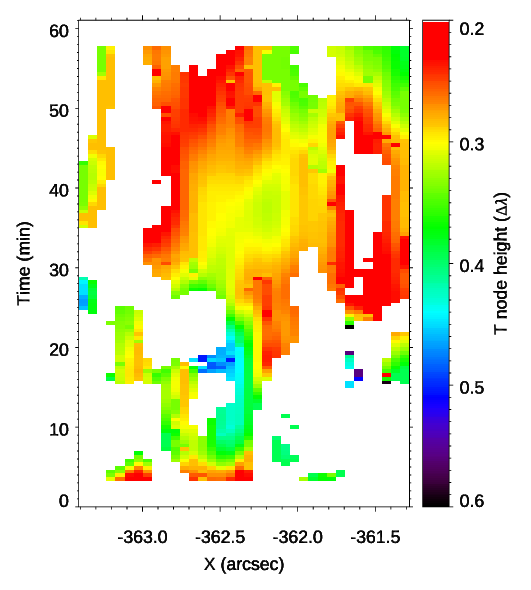}
\caption{Time-distance plot of the height of the temperature resonant node at 6.1 mHz along the location indicated by the horizontal black line in Fig. \ref{fig:nodos_maps}. In the white areas the node location was not identified.} \label{fig:nodos_time}
\end{figure}

\subsection{Temperature resonant nodes}

Figure \ref{fig:nodos_maps} shows maps of the the height of the temperature node at four selected time steps. Node locations are not identified in most of the field of view, especially out of the pore region. In contrast, we are able to quantify the node height in a significant area within the pore for each time step, indicating that active regions exhibit clear signatures of pure standing oscillations more often than quiet Sun regions. The region of coverage of pure standing oscillations changes with time, which means that at each time step we are able to determine the temperature node height of a different area of the pore. Also, the node height shows temporal fluctuations. Figure \ref{fig:nodos_time} illustrates a time-distance diagram of the temperature node height for the spatial locations along the black horizontal line plotted in Fig. \ref{fig:nodos_maps}. While the coverage is low in the first half of the temporal series, we can see how the $\Delta\lambda$ where the node is identified exhibits remarkable changes, both in space and time. At the location $X=-362.5\arcsec$, the temperature node changes from $\Delta\lambda\approx 0.5$ \AA\ at $t=18$ min to $\Delta\lambda\approx 0.2$ \AA\ at $t=53$ min. The co-existence of standing and propagating waves tends to produce a smoother transition in the oscillatory phase instead of a 180$^{\circ}$ jump, which results in an estimation of a deeper node. In fact, in the times around the $\Delta\lambda\approx 0.5$ \AA\ estimation, the node height is undetermined since the transition is not sharp enough. We consider that those cases with a temperature resonant node at very deep layers correspond to mixed standing and propagating waves, rather than striking shifts in the node height. 

We have constructed a map of the temperature node height at 6.1 mHz by combining all the time steps (inside the cone of influence at that frequency) into a single map with the median value of the node height (Fig. \ref{fig:nodos_average}). Within the pore region, the node height spans from $\Delta\lambda\approx 0.4$ \AA\ to $\Delta\lambda\approx 0.2$ \AA. The highest values are generally found in the central part of the pore, but also near the edges in the north half. Other regions of the pore exhibit an approximately constant height of $\Delta\lambda\approx 0.4$ \AA, with some small patches of a deeper estimated height probably produced by the contribution of the propagating waves to the oscillatory phase. No differences are found between the light bridge and other regions at the outer parts of the pore. Regarding the area surrounding the pore, the temperature nodes in the east region are mostly undetermined, and those locations where a height is assigned to the node show large values ($\Delta\lambda\ge 0.5$ \AA). As previously discussed, these measurements are found in those cases where the oscillations exhibit signatures of propagating waves superimposed to the standing oscillations. At the west side of the field of view, the node height is determined for most of the locations. The more inclined magnetic field in the regions surrounding the pore, and even at its edges, increases the path length of the waves to the transition region. As previously found, the signatures of standing oscillations decrease farther from the reflecting layer. Therefore, the increased field inclination, through the longer wave path, may contribute to the detection of propagating waves in the regions surrounding the pore. It may also affect the spatial variation of the node height. The outer regions of the pore tend to exhibit a deeper nodal layer, and the wave path at these locations could be even longer due to the greater inclination of the magnetic field.

\section{Discussion and conclusions}\label{sect:conclusions}

Despite being the first oscillatory phenomena ever reported in sunspots, three-minute umbral chromospheric oscillations are not yet fully understood, and many works are carried out every year to shed some light about this topic. Two main mechanisms have been proposed to explain this phenomenon. One of them is the existence of a chromospheric resonant cavity produced by sharp temperature gradients above the temperature minimum and the transition region \citep{Zhugzhda+Locans1981, Fleck+Deubner1989, Wood1990, YellesChaohuche+Abdelatif2005, Zhugzhda2008, Botha+etal2011}. The other mechanism is the upward propagation of waves with frequencies above the cutoff value \citep{Bel+Leroy1977, Zhugzhda+Dzhalilov1984c}. Numerous studies have found evidences of wave propagation from the photosphere to the chromosphere \citep[\eg,][]{Lites1984, Centeno+etal2006, Felipe+etal2010b, Kanoh+etal2016, KrishnaPrasad+etal2017}, apparently supporting the latter mechanism. However, recent works have also found indications of a resonant cavity \citep[see][]{Jess+etal2020, Felipe+etal2020, Felipe2021}, confirming that both mechanisms play a role in shaping the umbral chromospheric oscillations \citep{Felipe2019}.

Here, we report a strong evindence for the existence of standing oscillations in a pore chromosphere and, thus, of a chromospheric resonant cavity in solar active regions. We have found the sudden phase jumps of 180$^{\circ}$ and power dips that take place at the location of a resonant node. To the best of our understanding, this is the first time that this finding is reported. We have analyzed the phase of the velocity and temperature fluctuations at multiple heights probed by the \Ha\ line. The use of wavelet transforms allowed us to characterize the oscillations and their phase in time and frequency domains. 

At numerous time steps and locations within the pore we find pure standing oscillations, which are characterized by multiple insights. First, they exhibit a phase difference of $\pm90^{\circ}$ between velocity and temperature fluctuations \citep[\eg,][]{Deubner1974, Al+etal1998} at 6.1 mHz. This phase relation was also employed by \citet{Felipe+etal2020} to detect the existence of standing oscillations in the high chromosphere (observations in \HeI\ 10830 \AA) and upper photosphere (\NaIDtwo), and more recently by \citet{Sangal+etal2026} with \Ha\ data. Second, temperature fluctuations oscillate in phase at the highest layers sampled by the \Ha\ line, and also at the heights from $\Delta\lambda\approx0.4$ to $\Delta\lambda\approx0.6$ \AA\ (with the exact layers depending on the spatial location and time). In between, a 180$^{\circ}$ jump in the temperature phase and a remarkable dip in the temperature power is generally found around $\Delta\lambda\approx0.3$ \AA. This is clear evidence of the presence of a temperature resonant node for 6.1 mHz frequency waves at that atmospheric height. This phase jump is also detected in the phase differences between velocity and temperature fluctuations measured as a function of height. At all the layers above $\Delta\lambda=0.6$ \AA, the velocity is oscillating in phase. It also confirms the presence of standing waves but, contrary to the temperature, we do not find velocity resonant nodes at the atmospheric heights probed by \Ha. A velocity nodal plane must exist between the first-order temperature node that we have detected and the transition region \citep[][]{Fleck+Deubner1989, Felipe+etal2020}, but it must lie above the atmospheric layers probed by the \Ha\ line in this pore.

\citet{Sangal+etal2026} also reported standing oscillations from \Ha\ multi-height observations in a sunspot umbra. However, their results show that the intensity (temperature) oscillations are predominantly in-phase for all the probed $\Delta\lambda$. This may suggest some fundamental differences between pore and umbral atmospheres in the resonant cavity structure and/or the height response of the \Ha\ line. They also measured the phase differences between the intensity at several $\Delta\lambda$ and a single Doppler velocity signal sensitive to the chromosphere. They found some rapid transitions in the phase from $\Delta\lambda=0.6$ to $\Delta\lambda=0.4$ \AA. Although this transition may be related to the presence of a velocity resonant node, their results do not show a clear 180$^{\circ}$ jump and their intensity-velocity phase shifts are computed between oscillatory signals sampling different atmospheric heights, which prevented a definitive identification of a nodal layer.

\begin{figure}[ht]  
\centering
\includegraphics[width=0.45\textwidth]{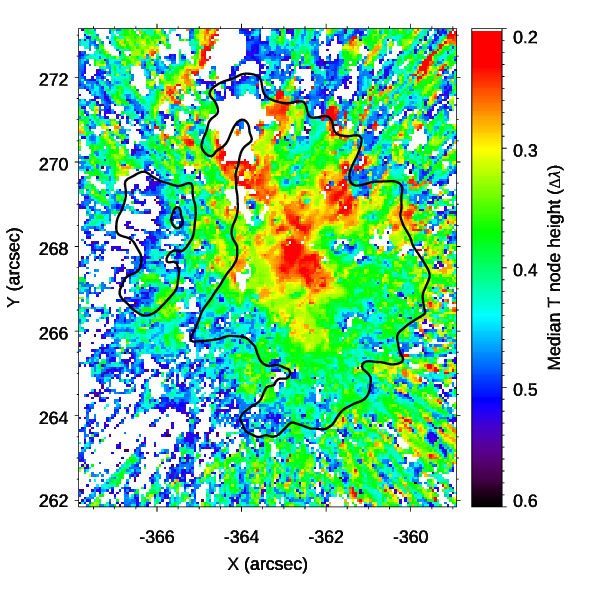}
\caption{Map of the temporal median of the temperature resonant node height at 6.1 mHz. In the white areas the node location was not identified at any time step. Black lines show contours of constant intensity in the average continuum intensity, delimiting the pore region.} \label{fig:nodos_average}
\end{figure}

The umbral chromosphere has been described as a leaky resonator by numerous observational studies \citep{Jess+etal2012b, Felipe+etal2018b, Sangal+etal2026} as well as numerical simulations \citep{Botha+etal2011, Snow+etal2015, Felipe+etal2020, Miriyala+etal2025}. At the transition region, reflection is not complete and waves can keep propagating into the corona. The reflected waves are trapped in an imperfect chromospheric cavity where partially standing oscillations are found, but they co-exist with upward or downward propagating waves. In this work, we also find evidence of mixed propagating and standing waves. They can be seen as departures of $\Delta\Phi_{\rm VI}$ from $\pm90^{\circ}$, as a smooth 180$^{\circ}$ transition in the phase around resonant nodes rather than sudden jumps, and as progressive changes in $\Delta\Phi_{\rm II}$ with height (bottom panels from Fig. \ref{fig:dphase_height}). The ratio of standing and propagating waves changes with the location and time. At a fixed location, the phase differences can show indications of pure standing oscillations during some times and then change to mostly propagating waves. We have also reported temporal variations in the height of the temperature nodal layer. \citet{Felipe+etal2025} identified short-term changes in the phase relations between the velocity inferred from NLTE inversions of the \CaII\ 8542 \AA\ line and that measured near the core of \Ha. These changes were interpreted as dynamic variations in the location of a velocity resonant node caused by changes in the height of the transition region. Other effects, such as local magnetoconvection, could also affect the cavity size and, consequently, the height of the resonant nodes. However, we do not find clear differences in the location of the nodes between the light bridge and the surrounding pore region. We therefore speculate that the main contributor to the temporal variations in node height is the dynamic variation in the height of the transition region.

Another open question is the depth of the resonant cavity. \citet{Felipe+etal2025} found propagating waves at the low chromosphere (at the formation height of the core of the \CaII\ 8542 \AA\ line) in the umbral regions with high amplitude oscillations. In contrast, other regions with weaker waves were consistent with standing oscillations at those heights. Recently, \citet{Chambers+etal2026} reported standing oscillations in the central part of a sunspot umbra at the upper photospheric layers probed by the \NaIDone\ and \NaIDtwo\ lines, in agreement with the previous finding from \citet{Felipe+etal2020}, who detected standing oscillations in \NaIDtwo\ data. Our results show that, even at those places and times where almost pure standing oscillations are found, waves are propagating at the heights probed by the wings of \Ha\ with $\Delta\lambda>0.65$ \AA. In other cases, wave propagation is found at higher layers. 

The detection of the resonant nodes opens a great opportunity to develop new seismic techniques to probe the umbral chromosphere, as proposed by \citet{Felipe+etal2020}. Figure \ref{fig:nodos_average} illustrates the spatial distribution of the height of the temperature resonant node at 6.1 mHz. They are located at distance $h$ from the height of the reflection layer given by \citep[\eg,][]{Fleck+Deubner1989}  

\begin{equation}
  h=\frac{n}{2}\frac{c_{\rm S}}{\nu},
\label{eq:nodos_T}
\end{equation}

\noindent where $n$ is the order of the node, $c_{\rm S}$ is the sound speed, and $\nu$ is the wave frequency. For $n=1$ and a chromospheric sound speed $c_{\rm S}=8$ km s$^{-1}$, the temperature nodes that we have detected for waves with $\nu=6.1$ mHz are 656 km below the height where those waves are reflected. This analysis can potentially provide the height of the seismic transition region where waves are reflected if the formation height of \Ha\ wings is characterized. However, we note that the understanding of \Ha\ formation is a formidable endeavor by itself \citep{Leenaarts+etal2012}, well beyond the scope of this paper. Waves with different frequencies are reflected at different transition region heights, with the reflection taking place at lower heights for lower frequency waves. This can be seen in the comparison between the location of the nodes obtained from a frequency-independent reflecting layer according to Eq. \ref{eq:nodos_T} and that inferred from numerical simulations in Fig. 1 from \citet{Felipe+etal2020}. This numerical modeling also shows that the frequency dependence of the seismic transition region changes with the sharpness of the temperature gradients, which could also be inferred from seismological methods. In the future, we plan to determine the location of the nodes at multiple frequencies, and to extend the analysis to a larger sample of active regions, including sunspots.

\begin{acknowledgements}
Financial support from grants PID2024-156538NB-I00, funded by MCIN/AEI/ 10.13039/501100011033 and by “ERDF A way of making Europe”, and from grant CNS2023-145233 funded by MICIU/AEI/10.13039/501100011033 and by “European Union NextGeneration EU/PRTR” is gratefully acknowledged. TF acknowledges grant RYC2020-030307-I funded by MCIN/AEI/ 10.13039/501100011033 and by “ESF Investing in your future”. This work is part of grant CEX2025-001609-S, awarded to the Instituto de Astrof\'isica de Canarias under the Severo Ochoa Centre of Excellence program and funded by MICIU/AEI/10.13039/501100011033. The Swedish 1-m Solar Telescope is operated on the island of La Palma by the Institute for Solar Physics of Stockholm University in the Spanish Observatorio del Roque de los Muchachos of the Instituto de Astrof\'isica de Canarias. The Swedish 1-m Solar Telescope, SST, is co-funded by the Swedish Research Council as a national research infrastructure (registration number 4.3-2021-00169). Wavelet software was provided by C. Torrence and G. Compo, and is available at URL: \url{http://atoc.colorado.edu/research/wavelets/}.

\end{acknowledgements}

\bibliographystyle{aa} 
\bibliography{biblio.bib}
\end{document}